\documentclass[aps, prd, amsmath, floats, floatfix, twocolumn, nofootinbib, superscriptaddress]{revtex4-1}
\usepackage{graphicx}
\usepackage{color}
\usepackage{latexsym}
\usepackage[colorlinks]{hyperref}

\newcommand{\be}{\begin{eqnarray}}
\newcommand{\ee}{\end{eqnarray}}

\begin{document}

\title{Towards precision tests of general relativity with black hole X-ray reflection spectroscopy}

\author{Ashutosh~Tripathi}
\affiliation{Center for Field Theory and Particle Physics and Department of Physics, Fudan University, 200438 Shanghai, China}

\author{Sourabh~Nampalliwar}
\affiliation{Theoretical Astrophysics, Eberhard-Karls Universit\"at T\"ubingen, 72076 T\"ubingen, Germany}

\author{Askar~B.~Abdikamalov}
\affiliation{Center for Field Theory and Particle Physics and Department of Physics, Fudan University, 200438 Shanghai, China}

\author{Dimitry~Ayzenberg}
\affiliation{Center for Field Theory and Particle Physics and Department of Physics, Fudan University, 200438 Shanghai, China}

\author{Cosimo~Bambi}
\email[Corresponding author: ]{bambi@fudan.edu.cn}
\affiliation{Center for Field Theory and Particle Physics and Department of Physics, Fudan University, 200438 Shanghai, China}

\author{Thomas~Dauser}
\affiliation{Remeis Observatory \& ECAP, Universit\"{a}t Erlangen-N\"{u}rnberg, 96049 Bamberg, Germany}

\author{Javier~A.~Garc{\'\i}a}
\affiliation{Cahill Center for Astronomy and Astrophysics, California Institute of Technology, Pasadena, CA 91125, USA}
\affiliation{Remeis Observatory \& ECAP, Universit\"{a}t Erlangen-N\"{u}rnberg, 96049 Bamberg, Germany}

\author{Andrea~Marinucci}
\affiliation{Dipartimento di Matematica e Fisica, Universit\'a degli Studi Roma Tre, 00146 Roma, Italy}

\begin{abstract}
Astrophysical black hole systems are the ideal laboratories for testing Einstein's theory of gravity in the strong field regime. We have recently developed a framework which uses the reflection spectrum of black hole systems to perform precision tests of general relativity by testing the Kerr black hole hypothesis. In this paper, we analyze \textsl{XMM-Newton} and \textsl{NuSTAR} observations of the supermassive black hole in the Seyfert~1 galaxy MCG--06--30--15 with our disk reflection model. We consider the Johannsen metric with the deformation parameters $\alpha_{13}$ and $\alpha_{22}$, which quantify deviations from the Kerr metric. For $\alpha_{22} = 0$, we obtain the black hole spin $0.928 < a_* < 0.983$ and $-0.44 < \alpha_{13} < 0.15$. For $\alpha_{13} = 0$, we obtain $0.885 < a_* < 0.987$ and $-0.12 < \alpha_{22} < 1.05$. The Kerr solution is recovered for $\alpha_{13} = \alpha_{22} = 0$. Thus, our results include the Kerr solution within statistical uncertainties. Systematic uncertainties are difficult to account for, and we discuss some issues in this regard.
\end{abstract}

\maketitle



\section{Introduction}

Einstein's gravity has been extensively tested in the weak field regime, its theoretical predictions being largely confirmed by experiments in the Solar System and radio observations of binary pulsars~\cite{will}. The strong field regime, on the other hand, is still largely unexplored. There are many alternative and modified theories of gravity that have the same predictions as Einstein's gravity for weak fields and present deviations only when gravity becomes strong. Astrophysical black holes give us an opportunity to test the predictions of Einstein's gravity in the strong field regime~\cite{r1,r2,r3,r4,r5}.

In 4-dimensional Einstein's gravity, the only stationary and asymptotically flat vacuum black hole solution, which is regular on and outside the event horizon, is the Kerr metric~\cite{h1,h2,h3}. The spacetime around astrophysical black holes is thought to be well approximated by this solution. Testing the Kerr black hole hypothesis with astrophysical black holes is thus a test of Einstein's gravity in the strong field regime, and can be seen as the counterpart of Solar System experiments aimed at verifying the Schwarzschild solution in order to test Einstein's gravity in the weak field regime~\cite{pp1,pp1b,pp2,pp3,pp3b,pp4,pp5,pp6,pp7}.

In this work, we study the X-ray spectrum of the supermassive black hole in MCG--06--30--15 with the reflection model {\sc relxill\_nk}~\cite{noi1} as a step in our program to test the Kerr black hole hypothesis from the reflection spectrum of the disk of accreting black holes~\cite{noi2,noi3,noi4,noi5,noi6}. MCG--06--30--15 is a very bright Seyfert~1 galaxy and it has been observed for many years by different X-ray missions. It is the source in which a relativistically blurred iron K$\alpha$ line was clearly detected for the first time~\cite{asca}, and is thus one of the natural candidates for tests of Einstein's gravity using X-ray reflection spectroscopy. We analyze simultaneous observations of \textsl{XMM-Newton}~\cite{r-xmm} and \textsl{NuSTAR}~\cite{r-nustar}, which provide both high energy resolution at the iron line (with \textsl{XMM-Newton}) and a broad energy band (with \textsl{NuSTAR}).

The contents of the paper are as follows. In Section~\ref{s-metric}, we briefly review the parameterized metric employed in our test and our previous results. In Sections~\ref{s-red} and \ref{s-ana} we present, respectively, our data reduction and analysis. Section~\ref{s-dis} is devoted to the discussion of our results and the conclusions. Throughout the paper, we adopt the convention $G_{\rm N} = c = 1$ and a metric with signature $(-+++)$.


\section{The reflection model {\sc relxill\_nk} \label{s-metric}}

{\sc relxill\_nk}~\cite{noi1} is the natural extension of {\sc relxill}~\cite{ref1,ref2} to non-Kerr spacetimes. It describes the disk's reflection spectrum of an accreting black hole~\cite{rev}. Here we employ the Johannsen metric. In Boyer-Lindquist-like coordinates, the line element reads~\cite{tj}
\be\label{eq-jm}
ds^2 &=&-\frac{\tilde{\Sigma}\left(\Delta-a^2A_2^2\sin^2\theta\right)}{B^2}dt^2
+\frac{\tilde{\Sigma}}{\Delta A_5}dr^2+\tilde{\Sigma}d\theta^2 \nonumber\\
&&-\frac{2a\left[\left(r^2+a^2\right)A_1A_2-\Delta\right]\tilde{\Sigma}\sin^2\theta}{B^2}dtd\phi \nonumber\\
&&+\frac{\left[\left(r^2+a^2\right)^2A_1^2-a^2\Delta\sin^2\theta\right]\tilde{\Sigma}\sin^2\theta}{B^2}d\phi^2
\ee
where $M$ is the black hole mass, $a = J/M$, $J$ is the black hole spin angular momentum, $\tilde{\Sigma} = \Sigma + f$, and
\be
&&  \Sigma = r^2 + a^2 \cos^2\theta \, , \qquad
\Delta = r^2 - 2 M r + a^2 \, , \nonumber\\
&& B = \left(r^2+a^2\right)A_1-a^2A_2\sin^2\theta \, .
\ee
The functions $A_1$, $A_2$, $A_5$, and $f$ are
\be
&& A_1 = 1 + \sum_{n=3}^{\infty}\alpha_{1n}\left(\frac{M}{r}\right)^n \, , \quad
A_2 = 1 + \sum_{n=2}^{\infty}\alpha_{2n}\left(\frac{M}{r}\right)^n \, , \nonumber\\
&& A_5 = 1 + \sum_{n=2}^{\infty}\alpha_{5n}\left(\frac{M}{r}\right)^n  \, , \quad
f = \sum^{\infty}_{n=3}\epsilon_n\frac{M^n}{r^{n-2}} \, .
\ee
The ``deformation parameters'' $\{ \alpha_{1n} \}$, $\{ \alpha_{2n} \}$, $\{ \alpha_{5n} \}$, and $\{ \epsilon_n \}$ are used to quantify possible deviations from the Kerr background. In what follows, we restrict our attention to the deformation parameters $\alpha_{13}$ and $\alpha_{22}$, since these two have the strongest impact on the reflection spectrum~\cite{noi1}. In our analysis, we leave either one of $\alpha_{13}$ and $\alpha_{22}$ free, setting the other to zero. All other deformation parameters are set identically to zero. In order to avoid spacetimes with pathological properties, we require $| a_* | \le 1$, where $a_* = a/M = J/M^2$ is the dimensionless spin parameter, and~\cite{noi3}
\be
\label{eq-constraints}
&& - \left(1 + \sqrt{1 - a_*^2} \right)^2 < \alpha_{22} < \frac{\left( 1 + \sqrt{1 - a^2_*} \right)^4}{a_*^2} \, , 
\nonumber\\
&& \alpha_{13} > - \frac{1}{2} \left( 1 + \sqrt{1 - a^2_*} \right)^4 \, .
\ee

From the analysis of the reflection spectrum of astrophysical black holes with {\sc relxill\_nk} we can constrain the deformation parameters $\alpha_{13}$ and $\alpha_{22}$ and check whether they are consistent with zero, as required by the Kerr black hole hypothesis. In Ref.~\cite{noi2}, we analyzed \textsl{XMM-Newton}, \textsl{NuSTAR}, and \textsl{Swift} data of the supermassive black hole in 1H0707--495 and we got the first constraint of $\alpha_{13}$ (see~\cite{shenzhen} for the same constraints with an updated version of {\sc relxill\_nk}). In Refs.~\cite{noi3} and \cite{noi6}, we analyzed \textsl{Suzaku} data of, respectively, the supermassive black hole in Ark~564 and Mrk~335, and we constrained the deformation parameters $\alpha_{13}$ and $\alpha_{22}$. In Refs.~\cite{noi4} and \cite{noi5}, we tested the Kerr nature of the stellar-mass black holes in GX~339--4 and GS~1354--645, respectively.

For these five sources, three supermassive black holes and two stellar mass black holes, we have found that the value of the deformation parameters is consistent with zero at least within a 90\% confidence level (and usually within 68\% confidence level). The most stringent constraints have been obtained from GS~1354--645, where the bounds on $a_*$, $\alpha_{13}$, and $\alpha_{22}$ are (99\% of confidence level for two relevant parameters)
\be\label{eq-a13-gs}
&& a_* > 0.975 \quad -0.34 < \alpha_{13} < 0.16 \quad (\text{for } \alpha_{22} = 0) \, , \\
&& a_* > 0.975 \quad -0.09 < \alpha_{22} < 0.42 \quad (\text{for } \alpha_{13} = 0) \, .
\label{eq-a22-gs}
\ee
Our results are thus consistent with the Kerr black hole hypothesis, as expected. However, these results were not obvious {\it a priori} considering the possible systematic effects of our model, which are not fully under control. This, in turn, might be interpreted as the fact that the systematic uncertainties are subdominant for the current precision of our tests.

\begin{table}
\centering
\begin{tabular}{ccc}
\hline\hline
$\hspace{0.0cm}$ Mission $\hspace{0.0cm}$ & $\hspace{0.0cm}$ Observation ID $\hspace{0.0cm}$ & $\hspace{0.0cm}$ Exposure (ks) $\hspace{0.0cm}$ \\
\hline
\textsl{NuSTAR} & 60001047002& 23 \\
& 60001047003 & 127 \\
& 60001047005 & 30 \\ 
\hline
\textsl{XMM-Newton} & 0693781201 & 134 \\
& 0693781301 & 134 \\
& 0693781401 & 49 \\
\hline\hline
\end{tabular}
\caption{List of the observations analyzed in this work. \label{t-obs}}
\end{table}

\begin{figure*}[t]
\vspace{-1.0cm}
\begin{center}
\includegraphics[width=16.0cm]{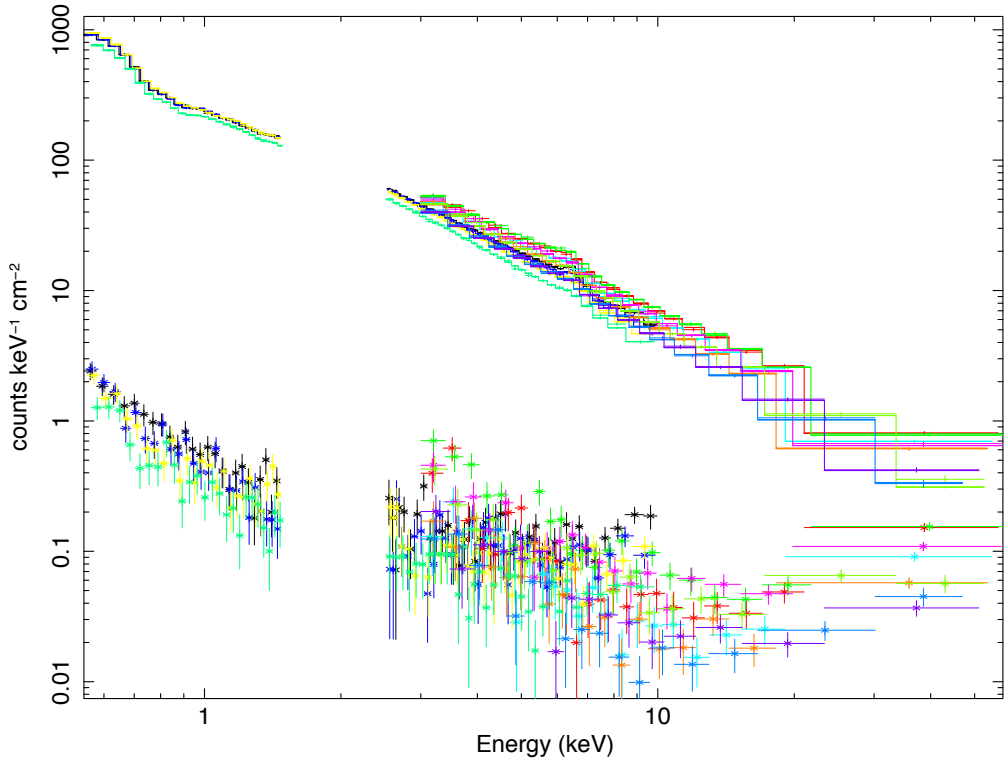}
\end{center}
\vspace{-1.2cm}
\caption{Source (the data in the upper part of the figure) and background (the data with the stars in the lower part of the figure) spectra for EPIC-Pn, FPMA, and FPMB for all the four flux states considered in this work. The data are divided by the response effective area of each particular channel. \label{f-back}}
\end{figure*}

\begin{figure}[t]
\vspace{-1.0cm}
\begin{center}
\includegraphics[width=9cm,trim={0.5cm 0 3cm 7cm},clip]{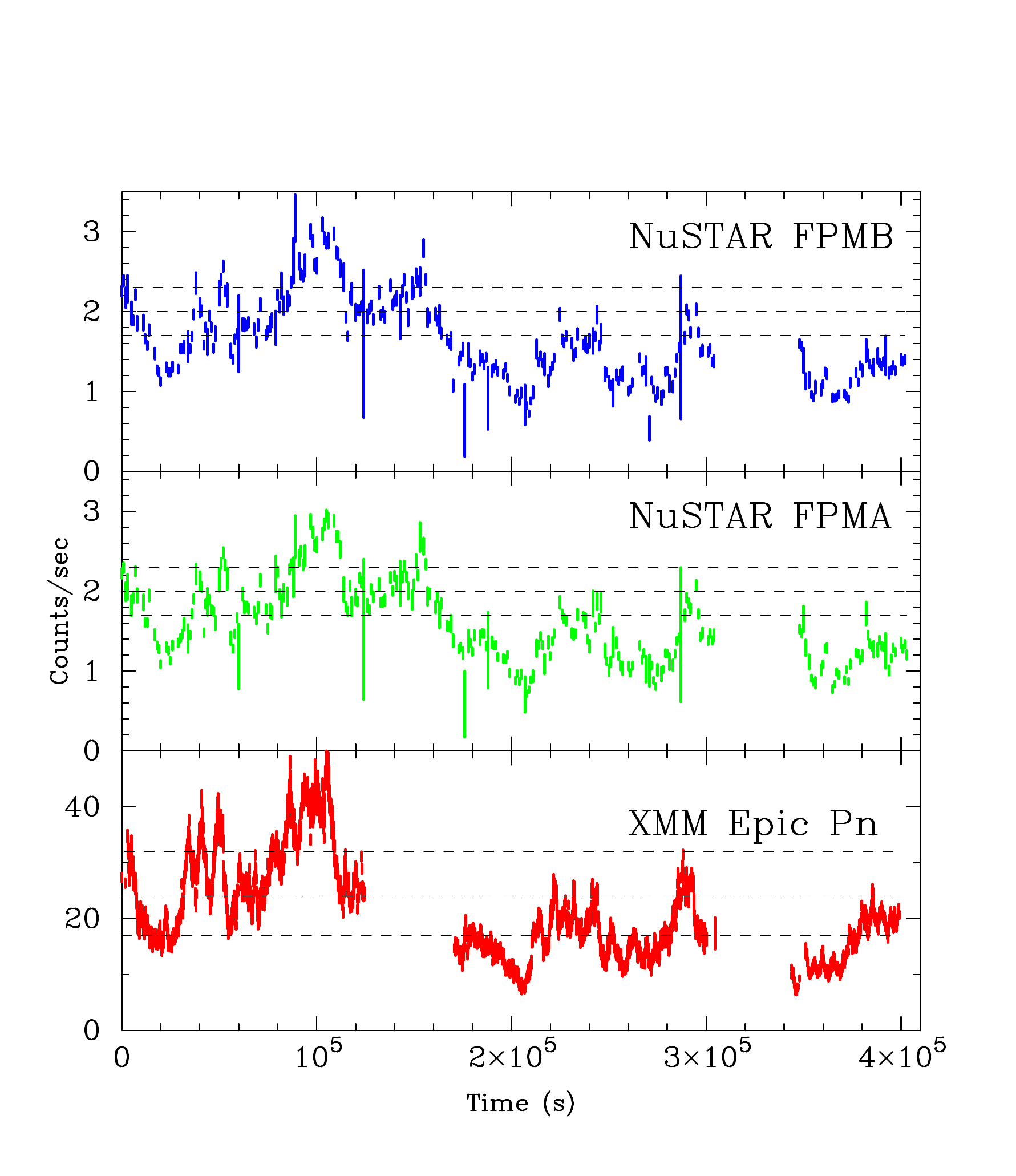}
\end{center}
\vspace{-0.7cm}
\caption{\textsl{NuSTAR}/FPMA, \textsl{NuSTAR}/FPMB and \textsl{XMM-Newton}/EPIC-Pn light curves. The three dashed horizontal lines separate the four different flux states.\label{f-lc}}
\end{figure}


\section{Observations and data reduction \label{s-red}}

MCG--06--30--15 is a very bright Seyfert~1 galaxy at redshift $z = 0.007749$ with many observations from different X-ray missions; see, for instance, Refs.~\cite{s1,s2,s3,s4,s5,s6,s7,s8,s9}. This source has a broad and very prominent iron K$\alpha$ line, so it is quite a natural candidate for our tests of the Kerr metric using X-ray reflection spectroscopy. However, the source is very variable, which requires some attention in the data analysis.

\textsl{NuSTAR} and \textsl{XMM-Newton} observed MCG--06--30--15 simultaneously starting on 29 January 2013 for a total time of $\sim$315~ks and $\sim$360~ks, respectively. Tab.~\ref{t-obs} shows the observation ID and their exposure time. A study of these data was reported in~\cite{s9}.

\textsl{NuSTAR} is comprised of two co-aligned telescopes with focal plane modules (FPMA and FPMB)~\cite{nustar}. The level~1 data products are analyzed using NuSTAR Data Analysis Software (NUSTARDAS). The downloaded raw data are converted into event files (level~2 products) using the HEASOFT task NUPIPELINE and using the latest calibration data files taken from NuSTAR calibration database (CALDB) version~20180312. The size of the source region is taken to be 70~arcsec centered at the source and that of the background is 100~arcsec taken from same CCD. The final products (light curves, spectra) are extracted using the event files and region files by running the NUPRODUCTS task. Spectra are rebinned to 70~counts per bin in order to apply $\chi^2$ statistics.

For \textsl{XMM-Newton}, observations from three consecutive revolutions are taken with the two EPIC cameras Pn and MOS1/2 operating in medium filter and small window mode~\cite{xmm}. Here, we are only using Pn data owing to their better quality~\cite{pn}. The MOS data are not used in our analysis because they suffered from high pile-up. SAS version~16.0.0 is used to convert the raw data into event files. These event files are then combined into a single fits file using ftool FMERGE. Good time intervals (GTIs) are generated using TABTIGEN and then used in filtering the event files. For source events, we take a circular region of 40~arcsec centered at the source. For background, we take a 50~arcsec region. After backscaling, response files are produced. Finally, in order to apply the $\chi^2$ statistics, spectra are rebinned in order to oversample the instrumental resolution by at least a factor of 3 and have 50~counts in each background-subtracted bin.

Source and background spectra of each instrument are shown in Fig.~\ref{f-back}.

As the source is highly variable, it is important to use simultaneous data so that variability is properly taken into account. We use ftool mgtime to find the common GTIs of the two telescopes.

\begin{table*}
\centering
\vspace{0.5cm}
\begin{tabular}{l|cccc|cccc}
\hline\hline
 & \multicolumn{4}{c}{Luminosity ($10^{43}$~erg~s$^{-1}$)} & \multicolumn{4}{c}{Flux ($10^{-10}$~erg~cm$^{-2}$~s$^{-1}$)} \\
\hline
Group & $\hspace{0.3cm}$ 1 $\hspace{0.3cm}$ & $\hspace{0.3cm}$ 2 $\hspace{0.3cm}$ & $\hspace{0.3cm}$ 3 $\hspace{0.3cm}$ & $\hspace{0.3cm}$ 4 $\hspace{0.3cm}$ & $\hspace{0.3cm}$ 1 $\hspace{0.3cm}$ & $\hspace{0.3cm}$ 2 $\hspace{0.3cm}$ & $\hspace{0.3cm}$ 3 $\hspace{0.3cm}$ & $\hspace{0.3cm}$ 4 $\hspace{0.3cm}$ \\
\hline
\textsl{XMM-Newton} & 0.59 & 0.83 & 1.12 & 1.54 & 0.44 & 0.62 & 0.84 & 1.15 \\
\textsl{NuSTAR}/FPMA & 1.01 & 1.24 & 1.50 & 2.03 & 0.76 & 0.93 & 1.12 & 1.53 \\
\textsl{NuSTAR}/FPMB & 1.04 & 1.27 & 1.52 & 2.07 & 0.78 & 0.95 & 1.14 & 1.55 \\
\hline\hline
\end{tabular}
\vspace{0.2cm}
\caption{Average luminosity (assuming $z=0.007749$) and average photon flux in the energy range 0.5-10~keV for \textsl{XMM-Newton} and 3-80~keV for \textsl{NuSTAR} for every flux state and instrument. Group~1 is for the low flux state, group~2 is for the medium flux state, group~3 is for the high flux state, and group~4 is for the very-high flux state. \label{t-flux}}
\end{table*}

\begin{table}
\centering
\vspace{0.5cm}
\begin{tabular}{l|cccc}
\hline\hline
Group & $\hspace{0.3cm}$ 1 $\hspace{0.3cm}$ & $\hspace{0.3cm}$ 2 $\hspace{0.3cm}$ & $\hspace{0.3cm}$ 3 $\hspace{0.3cm}$ & $\hspace{0.3cm}$ 4 $\hspace{0.3cm}$ \\
\hline
\textsl{XMM-Newton} & 1 & 1 & 1 & 1 \\
\textsl{NuSTAR}/FPMA & 1.060 & 1.044 & 1.043 & 1.053 \\
\textsl{NuSTAR}/FPMB & 1.089 & 1.065 & 1.058 & 1.072 \\
\hline\hline
\end{tabular}
\vspace{0.2cm}
\caption{Cross calibration constants between \textsl{XMM-Newton} and \textsl{NuSTAR}. The constant of \textsl{XMM-Newton} is frozen to 1. \label{t-cross}}
\end{table}

\begin{figure}[t]
\begin{center}
\includegraphics[width=9cm,trim={1cm 0 3cm 17cm},clip]{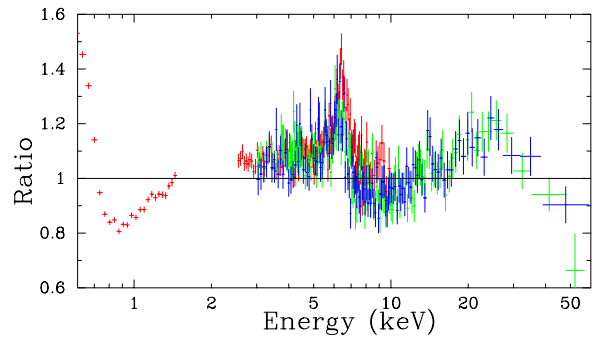}
\end{center}
\vspace{-0.7cm}
\caption{Data to best-fit model ratio for the model {\sc tbabs $\times$ cutoffpl} for the low flux state. We can clearly see the reflection features of the spectrum: broad iron line around 6~keV, Compton hump around 20~keV, and soft excess below 1~keV. Red crosses are used for \textsl{XMM-Newton}, green crosses for \textsl{NuSTAR}/FPMA, and blue crosses for \textsl{NuSTAR}/FPMB. \label{f-pow}}
\end{figure}

\begin{figure}[b]
\begin{center}
\includegraphics[width=9cm,trim={0.5cm 0 3cm 16cm},clip]{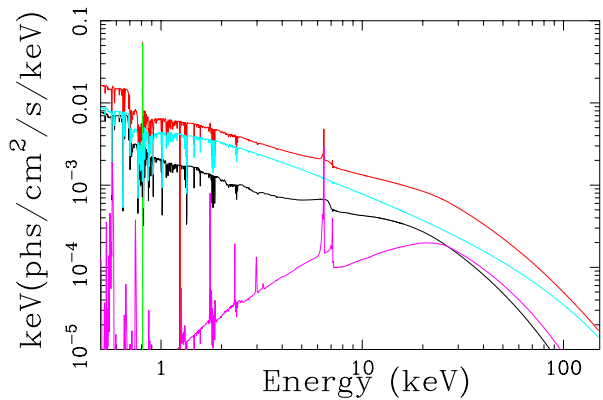}
\end{center}
\vspace{-0.9cm}
\caption{Spectrum of the best-fit of model $a$ of the low flux state (red) and its components: power law component (blue), relativistic reflection component (black), non-relativistic reflection component (magenta), and emission line (green). \label{f-model}}
\end{figure}

\begin{table}
\centering
\vspace{0.5cm}
\begin{tabular}{cccc}
\hline\hline
Model & $\hspace{0.3cm} \chi^2 \hspace{0.3cm}$ & $\hspace{0.3cm} \nu \hspace{0.3cm}$ & $\hspace{0.3cm} \chi^2/\nu \hspace{0.3cm}$ \\
\hline
0 & 47319 & 2727 & 17.35 \\
1& 18994 & 2726 & 6.9677 \\
2 & 11196 & 2718 & 4.1192 \\
3 & 10801 & 2711 & 3.9841 \\
4 & 10674 & 2702 & 3.9504 \\
5 & 3486.80 & 2691 & 1.29573 \\
6 & 3384.89 & 2689 & 1.25879 \\
7 & 3094.98 & 2688 & 1.15141 \\
8 & 3029.10 & 2685 & 1.12816 \\
\hline\hline
\end{tabular}
\vspace{0.2cm}
\caption{Statistics of the best-fit models. Model~0 is {\sc tbabs $\times$ cutoffpl}. In model~1, we add {\sc dustyabs} to model~0. In model~2, 3, and 4, we add, respectively, one, two, and three {\sc warmabs} to model~1. In model~5, we add {\sc relxill\_nk} to model~3. In model~6, we add {\sc xillver} to model~5. In model~7 and 8, we add one and two {\sc zgauss}, respectively, to model~6. \label{t-chi2}}
\end{table}


\begin{table*}
\centering
\vspace{0.5cm}
\begin{tabular}{l|cccc|cccc}
\hline\hline
Model & \multicolumn{4}{c}{$a$} & \multicolumn{4}{c}{$b$} \\
\hline
Group & 1 & 2 & 3 & 4 & 1 & 2 & 3 & 4 \\
\hline
{\sc tbabs} &&&& &&&& \\
$N_{\rm H} / 10^{22}$ cm$^{-2}$ & \multicolumn{4}{c}{$0.039^\star$} & \multicolumn{4}{c}{$0.039^\star$} \\
\hline
{\sc warmabs$_1$} &&&& \\
$N_{\rm H \, 1} / 10^{22}$ cm$^{-2}$ & $0.47^{+0.12}_{-0.06}$ & $1.163^{+0.015}_{-0.046}$ & $0.99^{+0.04}_{-0.03}$ & $0.25^{+0.04}_{-0.05}$ 
& $0.46^{+0.22}_{-0.06}$ & $1.16^{+0.04}_{-0.09}$ & $0.99^{+0.04}_{-0.11}$ & $0.25^{+0.09}_{-0.17}$ \\
$\log\xi_1$ & $1.86^{+0.04}_{-0.04}$ & $1.955^{+0.011}_{-0.020}$ & $1.922^{+0.014}_{-0.024}$ & $2.48^{+0.09}_{-0.13}$ 
& $1.86^{+0.08}_{-0.10}$ & $1.95^{+0.05}_{-0.06}$ & $1.92^{+0.03}_{-0.05}$ & $2.48^{+0.32}_{-0.18}$ \\
\hline
{\sc warmabs$_2$} &&&& \\
$N_{\rm H \, 2} / 10^{22}$ cm$^{-2}$ & $0.63^{+0.05}_{-0.06}$ & $0.02^{+0.02}_{-0.02}$ & $0.54^{+0.17}_{-0.11}$ & $0.72^{+0.10}_{-0.04}$ 
& $0.64^{+0.10}_{-0.31}$ & $0.02^{+0.02}_{-0.02}$ & $0.54^{+0.60}_{-0.23}$ & $0.72^{+0.20}_{-0.13}$ \\
$\log\xi_2$ & $1.904^{+0.024}_{-0.073}$ & $3.09^{+0.09}_{-1.16}$ & $3.23^{+0.05}_{-0.06}$ & $1.829^{+0.011}_{-0.020}$ 
& $1.90^{+0.11}_{-0.14}$ & $3.1_{-1.0}$ & $3.23^{+0.13}_{-0.17}$ & $1.83^{+0.13}_{-0.04}$ \\
\hline
{\sc dustyabs} &&&& \\
$\log \big( N_{\rm Fe} / 10^{21}$ cm$^{-2} \big)$ & \multicolumn{4}{c}{$17.411^{+0.006}_{-0.018}$} & \multicolumn{4}{c}{$17.41^{+0.04}_{-0.05}$} \\
\hline
{\sc cutoffpl} &&&& \\
$\Gamma$ & $1.952^{+0.007}_{-0.003}$ & $1.971^{+0.006}_{-0.010}$ & $2.010^{+0.004}_{-0.011}$ & $2.024^{+0.005}_{-0.011}$ & $1.952^{+0.018}_{-0.010}$ & $1.971^{+0.028}_{-0.013}$ & $2.010^{+0.030}_{-0.019}$ & $2.024^{+0.025}_{-0.039}$ \\
$E_{\rm cut}$ [keV] & $198^{+11}_{-26}$ & $157^{+20}_{-17}$ & $166^{+22}_{-23}$ & $278^{+116}_{-44}$
& $198^{+80}_{-50}$ & $157^{+71}_{-44}$ & $166^{+82}_{-38}$ & $189^{+86}_{-87}$ \\
$N_\text{\sc cutoffpl}$~$(10^{-3})$ & $8.29^{+0.10}_{-0.34}$ & $11.94^{+0.20}_{-0.27}$ & $14.30^{+0.25}_{-0.33}$ & $20.1^{+1.9}_{-1.9}$ & $8.3^{+0.6}_{-1.2}$ & $12.0^{+2.2}_{-1.1}$ & $14.3^{+1.1}_{-0.7}$ & $20.1^{+0.5}_{-2.3}$ \\ 
\hline
{\sc relxill\_nk} &&&& \\
$q_{\rm in}$ & $6.2^{+1.1}_{-1.0}$ & $7.0^{+0.6}_{-0.6}$ & $7.68^{+0.36}_{-0.21}$ & $8.07^{+0.50}_{-0.17}$
& $\sim 6$ & $7.0^{+2.8}_{-2.2}$ & $7.7^{+0.9}_{-1.1}$ & $8.1^{+0.5}_{-2.4}$ \\
$q_{\rm out}$ & \multicolumn{4}{c}{$3^\star$} & \multicolumn{4}{c}{$3^\star$} \\
$R_{\rm br}$ [$M$] & $2.88^{+0.04}_{-0.06}$ & $2.98^{+0.14}_{-0.15}$ & $3.28^{+0.12}_{-0.06}$ & $3.38^{+0.14}_{-0.35}$
& $2.88^{+0.17}_{-0.46}$ & $3.0^{+0.4}_{-0.9}$ & $3.28^{+0.13}_{-0.17}$ & $3.4^{+0.8}_{-0.6}$ \\
$i$ [deg] & \multicolumn{4}{c}{$31.4^{+1.3}_{-1.4}$} & \multicolumn{4}{c}{$31.5^{+2.6}_{-2.9}$} \\
$a_*$ & \multicolumn{4}{c}{$0.967^{+0.007}_{-0.013}$} & \multicolumn{4}{c}{$0.967^{+0.004}_{-0.056}$} \\
$\alpha_{13}$ & \multicolumn{4}{c}{$0.00^{+0.07}_{-0.20}$} & \multicolumn{4}{c}{$0^\star$} \\
$\alpha_{22}$ & \multicolumn{4}{c}{$0^\star$} & \multicolumn{4}{c}{$0.0^{+0.6}_{-0.1}$} \\
$z$ & \multicolumn{4}{c}{$0.007749^\star$} & \multicolumn{4}{c}{$0.007749^\star$} \\
$\log\xi$ & $2.88^{+0.04}_{-0.06}$ & $3.008^{+0.007}_{-0.047}$ & $3.064^{+0.020}_{-0.020}$ & $3.133^{+0.014}_{-0.021}$ 
& $2.88^{+0.07}_{-0.09}$ & $3.008^{+0.025}_{-0.063}$ & $3.064^{+0.017}_{-0.022}$ & $3.13^{+0.04}_{-0.05}$ \\
$A_{\rm Fe}$ & \multicolumn{4}{c}{$2.97^{+0.22}_{-0.14}$} & \multicolumn{4}{c}{$2.98^{+0.49}_{-0.28}$} \\
$N_\text{\sc relxill\_nk}$~$(10^{-3})$ & $0.125^{+0.012}_{-0.008}$ & $0.165^{+0.009}_{-0.007}$ & $0.299^{+0.009}_{-0.040}$ & $0.383^{+0.010}_{-0.061}$ & $0.125^{+0.012}_{-0.024}$ & $0.165^{+0.013}_{-0.021}$ & $0.300^{+0.158}_{-0.024}$ & $0.383^{+0.068}_{-0.015}$ \\ 
\hline
{\sc xillver} &&&& \\
$\log\xi'$ & \multicolumn{4}{c}{$0^\star$} & \multicolumn{4}{c}{$0^\star$} \\
$N_\text{\sc xillver}$~$(10^{-3})$ & \multicolumn{4}{c}{$0.058^{+0.004}_{-0.004}$} & \multicolumn{4}{c}{$0.058^{+0.011}_{-0.010}$} \\
\hline
{\sc zgauss} &&&& &&&&\\
$E_{\rm line}$ [keV] & \multicolumn{4}{c}{$0.8143^{+0.0008}_{-0.0032}$} & \multicolumn{4}{c}{$0.814^{+0.003}_{-0.003}$} \\
\hline
{\sc zgauss} &&&& &&&&\\
$E_{\rm line}$ [keV] & \multicolumn{4}{c}{$1.226^{+0.011}_{-0.008}$} & \multicolumn{4}{c}{$1.225^{+0.021}_{-0.020}$} \\
\hline
$\chi^2$/dof & \multicolumn{4}{c}{$3029.10/2685 = 1.12816$} & \multicolumn{4}{c}{$3029.15/2685 = 1.12818$} \\
\hline\hline
\end{tabular}
\vspace{0.2cm}
\caption{Summary of the best-fit values for model~$a$ ($\alpha_{13}$ free and $\alpha_{22} = 0$) and model~$b$ ($\alpha_{13} = 0$ and $\alpha_{22}$ free). The ionization parameter $\xi$ is in units erg~cm~s$^{-1}$. The reported uncertainties correspond to the 90\% confidence level for one relevant parameter. $^\star$ indicates that the parameter is frozen. See the text for more details. \label{t-fit}}
\end{table*}

\section{Spectral analysis \label{s-ana}}

MCG--06--30--15 is highly variable in the X-ray band. This could be due to passing clouds near the black hole along our line of sight and/or variations of the coronal geometry, as both phenomena can have a timescale shorter than our observations. To take such a source variability into account, we have arranged our data in four groups according to the flux state of the source (low flux state, medium flux state, high flux state, and very-high flux state). Since we have data from three instruments (\textsl{XMM-Newton}, \textsl{NuSTAR}/FPMA, and \textsl{NuSTAR}/FPMB), in the end we have to deal with 12~data sets. The data are divided into four flux states such that spectral data counts will be similar in each flux state as shown in Fig.~\ref{f-lc}. Luminosities and fluxes in the energy range 0.5-10~keV for \textsl{XMM-Newton} and 3-80~keV for \textsl{NuSTAR} for every data set are shown in Tab.~\ref{t-flux}. Note that our grouping scheme is different from that employed in Ref.~\cite{s9}, which was based on the hardness of the source.

To combine the \textsl{XMM-Newton} and \textsl{NuSTAR} data, we set the constant of \textsl{XMM-Newton} to 1 and we leave the constants of \textsl{NuSTAR}/FPMA and \textsl{NuSTAR}/FPMB free. After the fit, we check that the ratio between the constants of \textsl{NuSTAR}/FPMA and \textsl{NuSTAR}/FPMB is between 0.95 and 1.05. Tab.~\ref{t-cross} shows the values of these constants for every flux state.

As discussed in the appendix of Ref.~\cite{s9}, in the \textsl{XMM-Newton} data we see a spurious Gaussian around 2~keV. This is interpreted as an effect of the golden edge in the response file due to mis-calibration in the long-term charge transfer inefficiency (CTI), i.e.~how photon energies are reconstructed after detection. We solve this issue by simply ignoring the energy range 1.5-2.5~keV in the \textsl{XMM-Newton}/EPIC-Pn data. Such a region is not crucial for testing the Kerr metric and therefore its omission is not so important for the final result. We cannot add an {\it ad hoc} Gaussian to fit this feature because this would also modify the way in which the warm absorbers/ionized reflectors reproduce the data.

We first try to fit the data of the low flux state with a simple power law to identify the spectral features. The notable features above 3~keV are the iron K$\alpha$ line around 6.4~keV and the Compton hump at 20-30~keV (see Fig.~\ref{f-pow})~\cite{george91,ross05}. Below 3~keV, there are features from complex ionized absorbers~\cite{lee01,sako03}.  In~\cite{lee00}, the authors studied the low energy spectrum of this source and found that fitting requires two warm absorbers and one neutral absorber; this is the choice extensively adopted in the literature~\cite{s6,s9}. In order to fit the spectrum, we employ the model consisting of the following components: non-relativistic reflection from distant cold material, relativistic reflection from the ionized accretion disk and power law for primary emission. We use {\sc xillver} for the cold reflection~\cite{ref1}, {\sc relxill\_nk} for the blurred reflection~\cite{noi1}, and {\sc cutoffpl} for the power law emission with free cut-off energy. A narrow emission line and a narrow absorption line are also required. The combination of the above-mentioned models is convolved with two ionized absorbers, one dusty absorber, and Galactic absorption as mentioned in the literature~\cite{lee00,s6,s9}. Tab.~\ref{t-chi2} shows the improvement of the fit as we add new components to the model.

The final XSPEC model is {\sc tbabs $\times$ warmabs$_1$ $\times$ warmabs$_2$ $\times$ dustyabs $\times$ (cutoffpl + relxill\_nk + xillver + zgauss + zgauss)}. {\sc tbabs} describes the Galactic absorption and we set the column density $N_{\rm H} = 3.9 \cdot 10^{20}$~cm$^{-2}$~\cite{dickey}. {\sc warmabs$_1$} and {\sc warmabs$_2$} are two ionized absorbers and their tables are generated with {\sc xstar} v~2.41. {\sc dustyabs} is a neutral absorber which modifies the soft X-ray band due to the presence of dust around the source~\cite{lee00}. {\sc cutoffpl} is a power law with an exponential cut-off and describes the direct radiation from the Comptonized corona. {\sc relxill\_nk} is our disk's reflection model for the Johannsen spacetime~\cite{noi1}, where the reflection fraction parameter is set to $-1$, so there is no power-law from the corona because we prefer to use {\sc cutoffpl}. {\sc xillver} is the reflection spectrum from some ionized non-relativistic matter at larger distance~\cite{ref3}. After fitting the data with all the above mentioned model components, there are features at low energies that can be fit with gaussian profiles. One of the two {\sc zgauss} describes a narrow oxygen line around 0.81~keV and the other one describes a narrow absorption at 1.22~keV. The latter can be interpreted in terms of blueshifted oxygen absorption due to the presence of relativistic outflows~\cite{lei}. The spectrum of the best-fit model with its components for the low flux state is shown in Fig~\ref{f-model}.

\begin{figure*}[t]
\vspace{-1.0cm}
\begin{center}
\includegraphics[width=8.0cm,trim={2.2cm 0 3.2cm 15cm},clip]{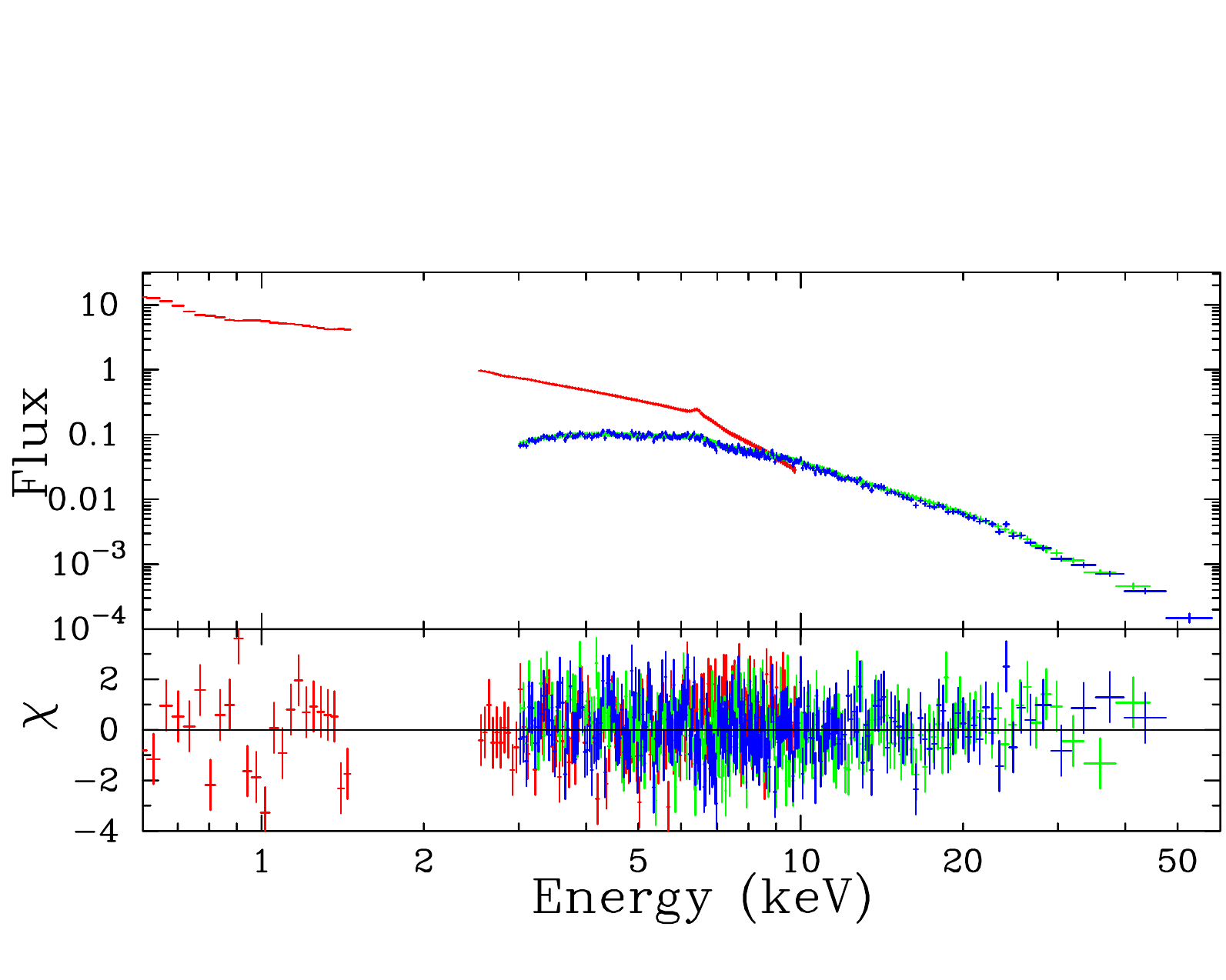}
\hspace{0.7cm}
\includegraphics[width=8.0cm,trim={2.2cm 0 3.2cm 15cm},clip]{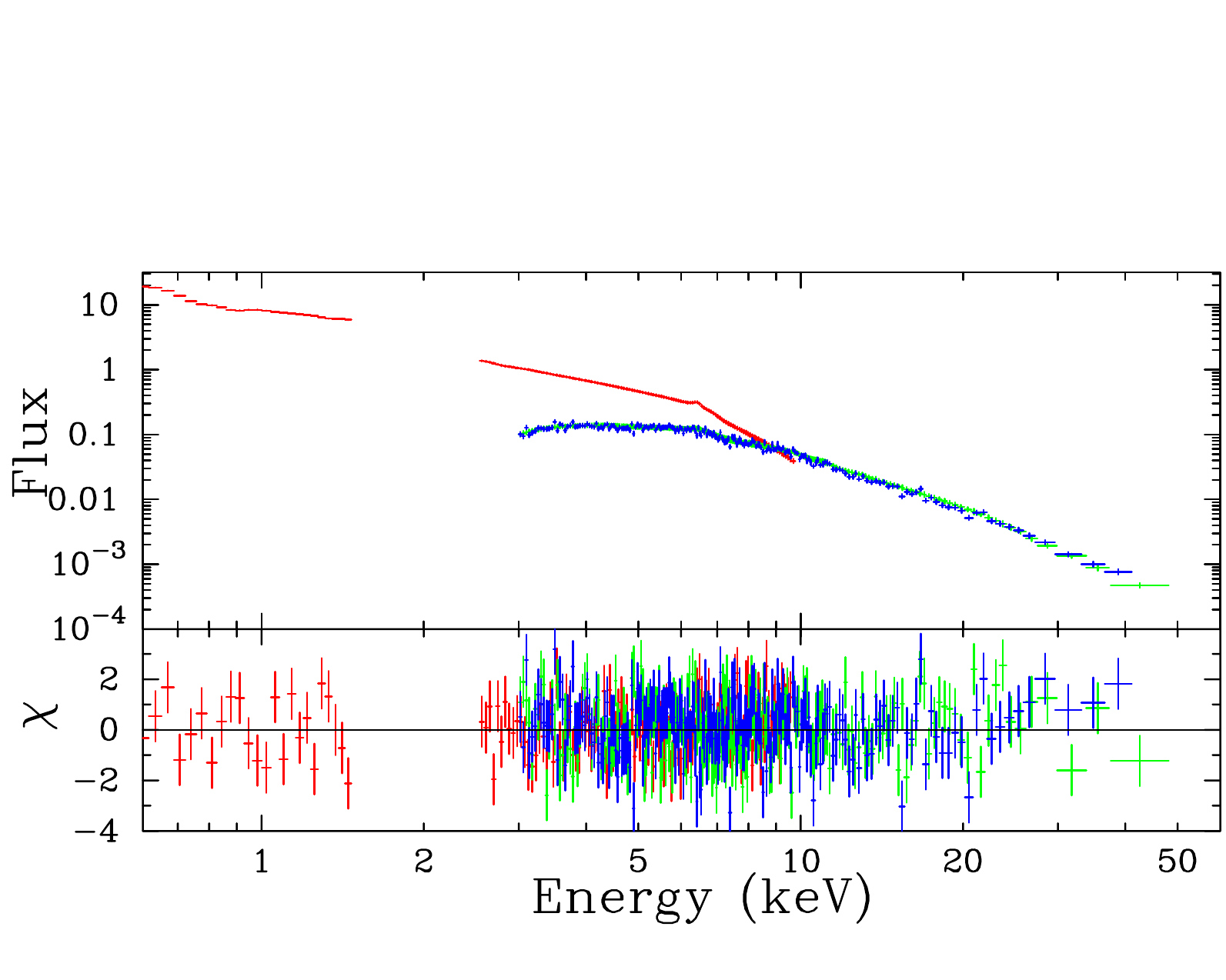} \\ 
\includegraphics[width=8.0cm,trim={2.2cm 0 3.2cm 18cm},clip]{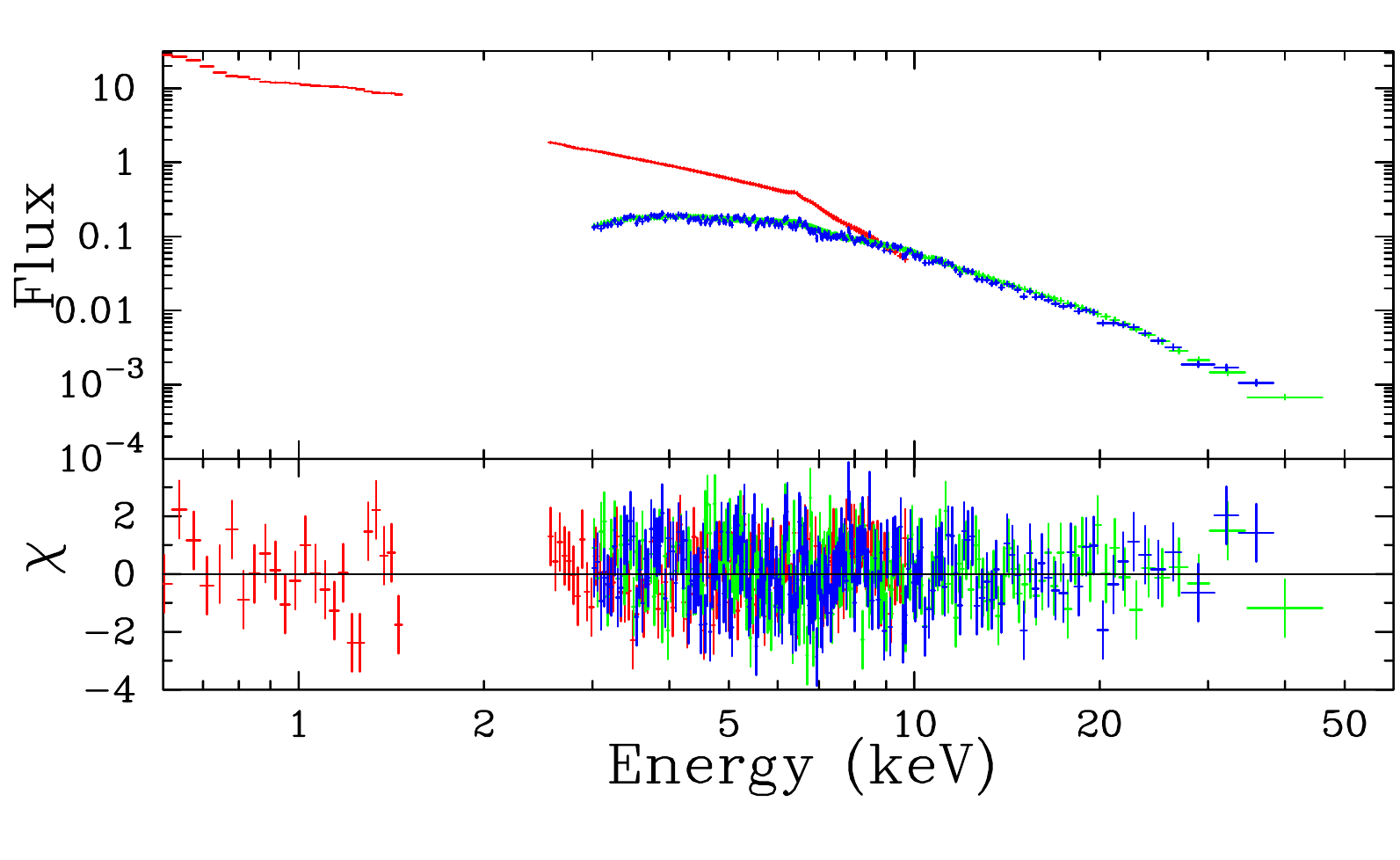}
\hspace{0.7cm}
\includegraphics[width=8.0cm,trim={2.2cm 0 3.2cm 18cm},clip]{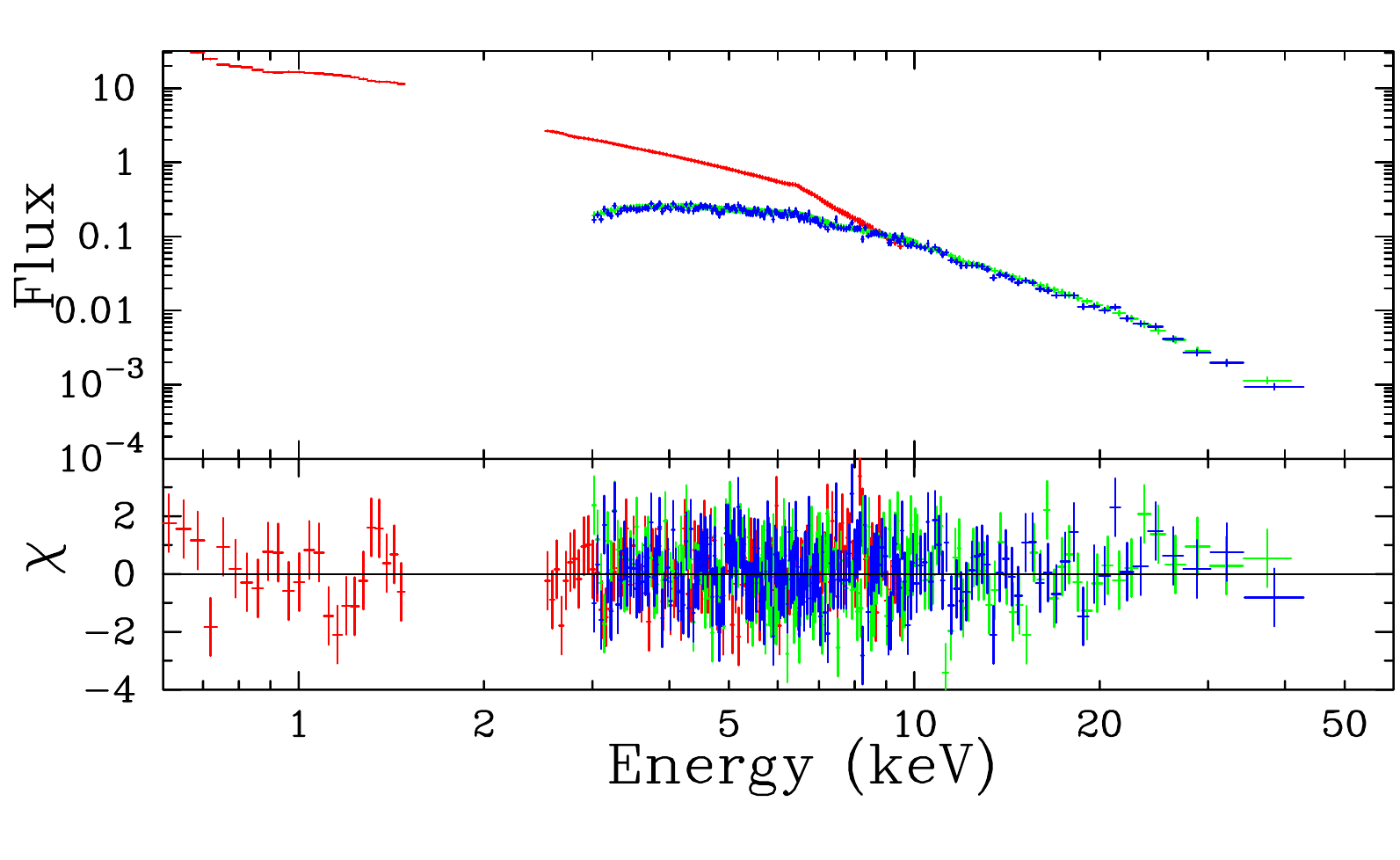}
\end{center}
\vspace{-0.7cm}
\caption{Best-fit model and standard deviations for model~$a$. The top left panel is for the low flux state, the top right panel is for the medium flux state, the bottom left panel is for the high flux state, and the bottom right panel is for the very-high flux state. Red crosses are used for \textsl{XMM-Newton}, green crosses for \textsl{NuSTAR}/FPMA, and blue crosses for \textsl{NuSTAR}/FPMB. \label{f-ratio}}
\end{figure*}

We have two models: model~$a$ in which $\alpha_{13}$ is free and $\alpha_{22} = 0$, and model~$b$ where $\alpha_{13} = 0$ and $\alpha_{22}$ can vary. Thus, we test for one non-zero deformation parameter at a time. The best-fit values are reported in Tab.~\ref{t-fit} for both model~$a$ and model~$b$. The estimated error is the 90\% confidence interval for one parameter of interest ($\Delta \chi$=2.71). Fig.~\ref{f-ratio} shows the quality of our fits with the residuals for model~$a$ (for model~$b$ we obtain very similar results).

The column densities and the ionization parameters of the two warm absorbers are allowed to vary from different flux states. The neutral iron density in {\sc dustyabs} is instead kept constant: it describes the absorption of the dust surrounding the source and its iron density should not change much among different flux states; see~\cite{s3} for more details about dust absorption in MCG--06--30--15. The photon index $\Gamma$ and the energy cut-off $E_{\rm cut}$ to describe the spectrum of the corona are allowed to vary with the flux state because the geometry of the corona can change over the observational timescale.

For the disk's reflection spectrum described by {\sc relxill\_nk}, we start with an emissivity profile described by a broken power-law and the inner emissivity index $q_{\rm in}$, the outer emissivity index $q_{\rm out}$, and the breaking radius $R_{\rm br}$ all free and allowed to vary with the flux state. However, we find that $q_{\rm out}$ is always consistent with 3, as we could expect in the case of a lamppost corona. For example, for model~$a$ we get $2.90^{+0.13}_{-0.12}$, $2.93^{+0.27}_{-0.19}$, $3.09^{+0.21}_{-0.16}$, and $2.78^{+0.28}_{-0.27}$ for the four flux states. We thus repeat the fit freezing $q_{\rm out}$ to 3. The inclination angle of the accretion disk, the black hole spin, the deformation parameters, and the iron abundances are clearly constant over the observation period. The ionization parameter $\xi$ is allowed to vary because it is affected by the geometry of the corona, which may change at different flux states. The normalization also varies among different flux states. We freeze the reflection fraction of {\sc relxill\_nk} to be $-1$ so that it only returns the reflection component. The power law component is modeled with {\sc cutoffpl} and the cut-off energy is left free because it can be estimated from the \textsl{NuSTAR} data.

In {\sc xillver}, the parameters are tied to those in {\sc relxill\_nk}, with the exception of the ionization and iron abundance. For the ionization parameter, we set $\log\xi' = 0$, as the non-relativistic reflection component is thought to be produced far away the black hole, in the outer part of the accretion disk or the molecular torus. The iron abundance is fixed at solar value as the distant cold reflector likely has a low iron abundance. The normalization is tied between different flux states as the distant reflector is not expected to vary much.

\begin{figure*}[t]
\begin{center}
\vspace{0.3cm}
\includegraphics[type=pdf,ext=.pdf,read=.pdf,width=0.5\textwidth]{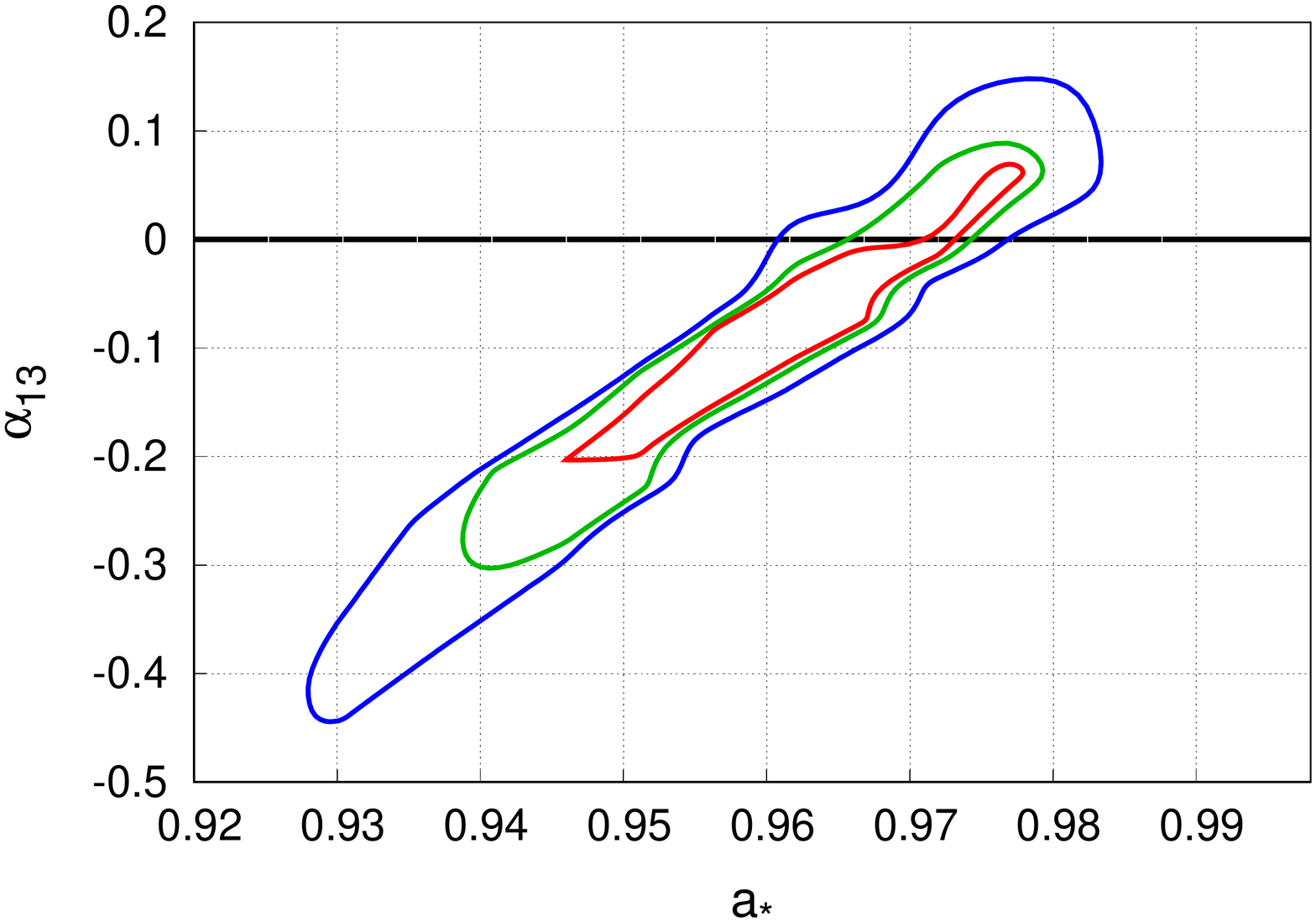}
\hspace{-0.5cm}
\includegraphics[type=pdf,ext=.pdf,read=.pdf,width=0.5\textwidth]{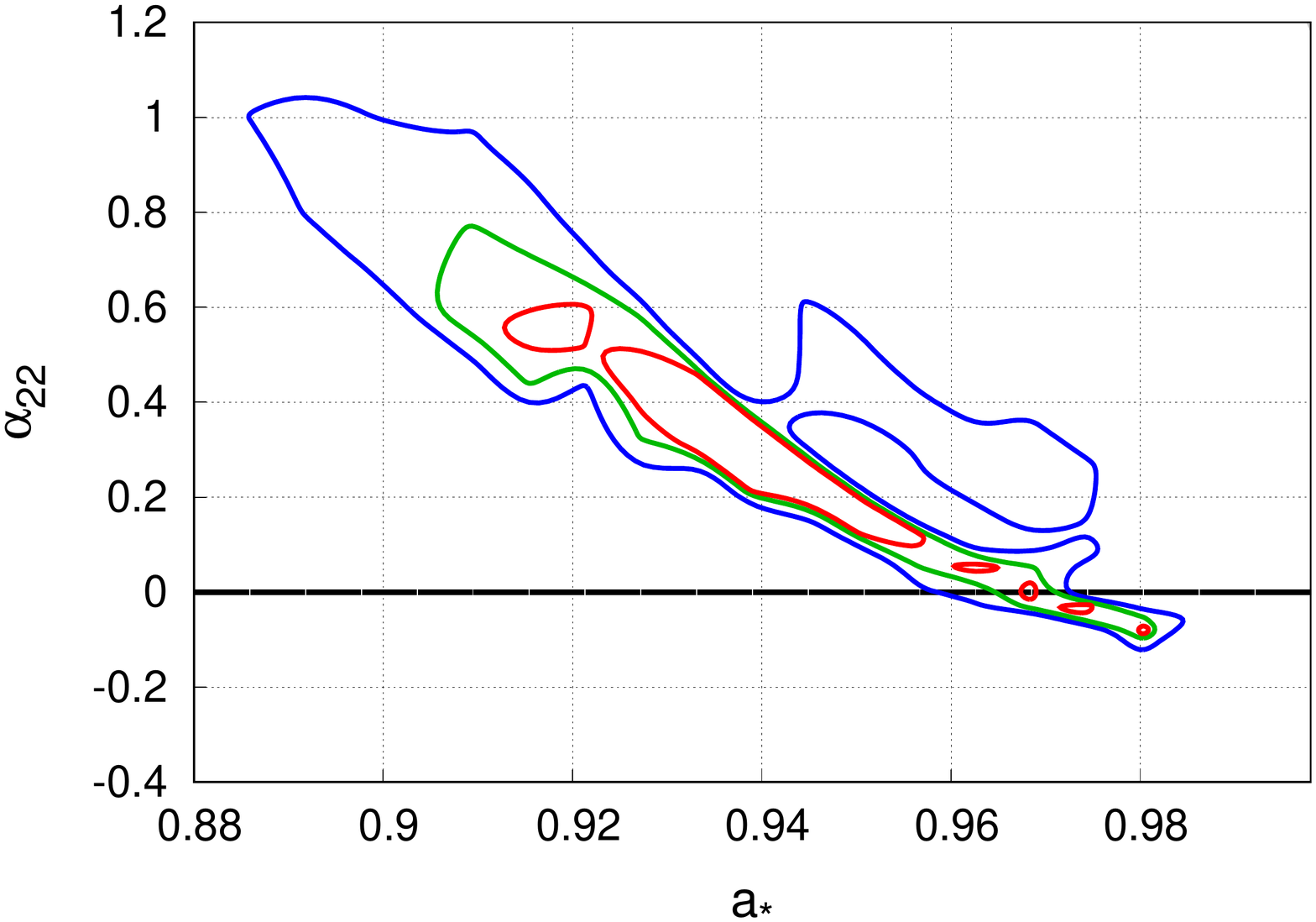}
\end{center}
\vspace{-1.3cm}
\caption{Constraints on the spin parameter $a_*$ and on the Johannsen deformation parameter $\alpha_{13}$ (left panel) and $\alpha_{22}$ (right panel). The red, green, and blue lines indicate, respectively, the 68\%, 90\%, and 99\% confidence level contours for two relevant parameters. The thick black horizontal line marks the Kerr solution. \label{f-plots}}
\end{figure*}

\begin{figure*}[h]
\begin{center}
\vspace{0.3cm}
\includegraphics[type=pdf,ext=.pdf,read=.pdf,width=0.5\textwidth]{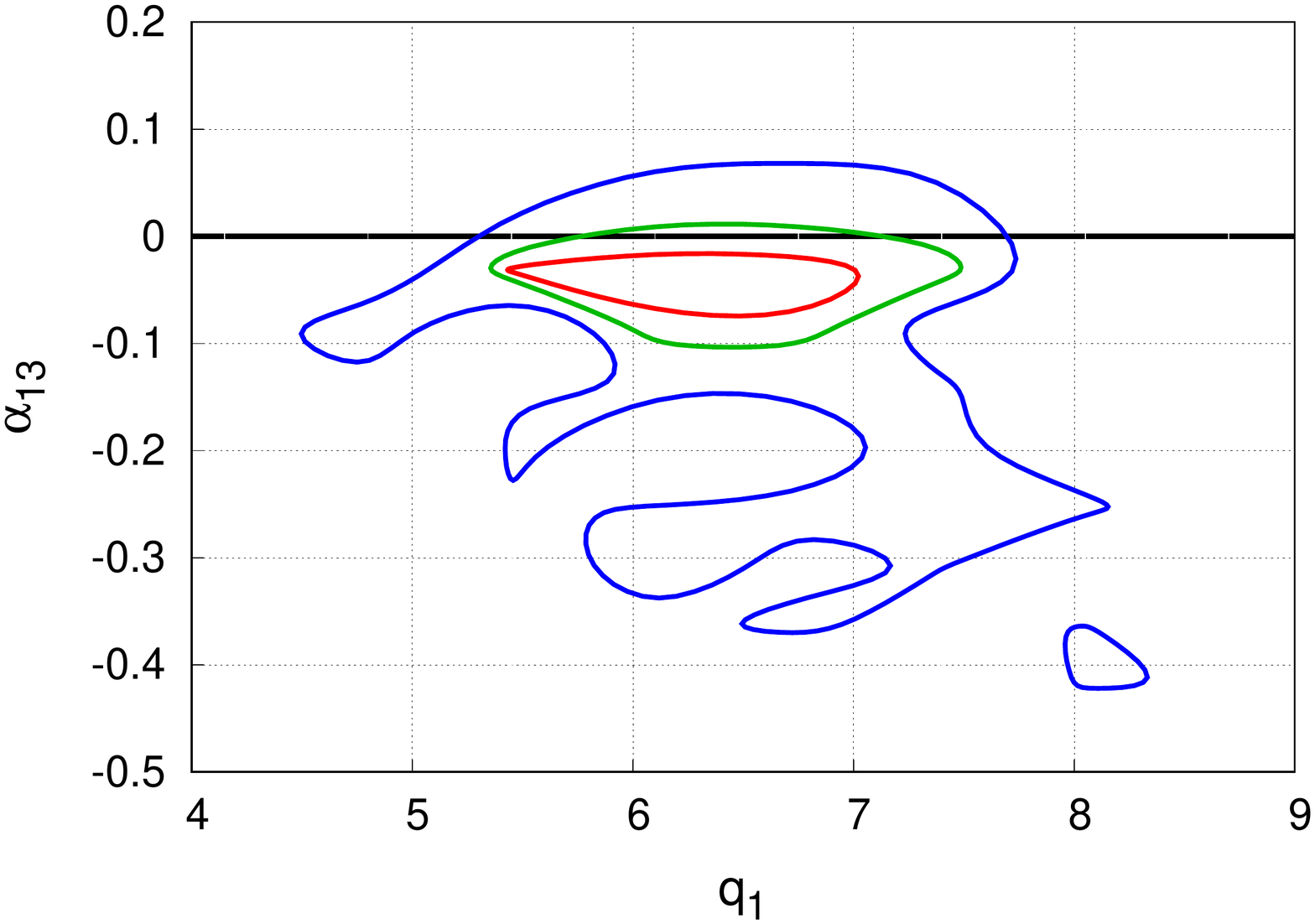}
\hspace{-0.5cm}
\includegraphics[type=pdf,ext=.pdf,read=.pdf,width=0.5\textwidth]{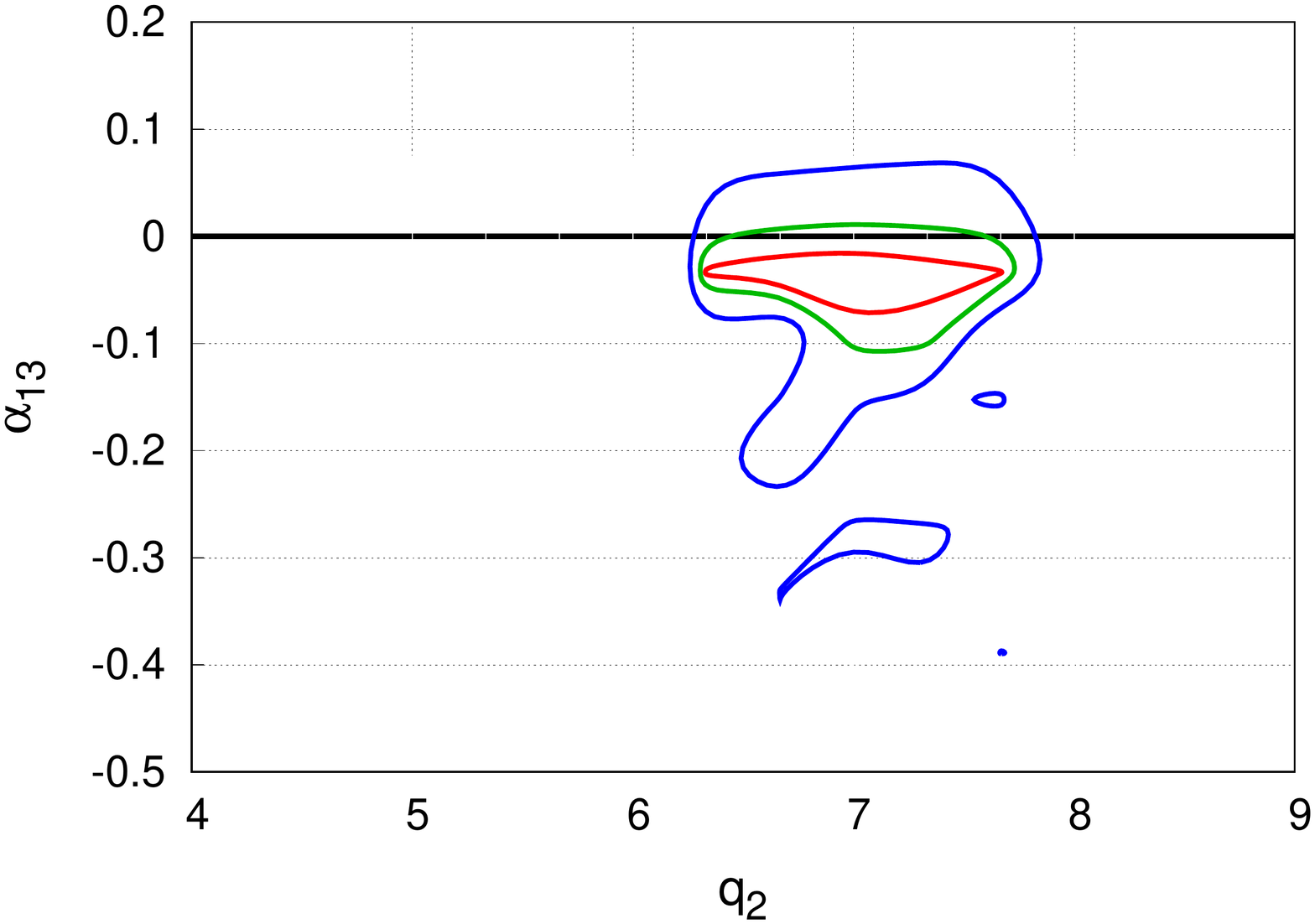} \\
\vspace{-1.0cm}
\includegraphics[type=pdf,ext=.pdf,read=.pdf,width=0.5\textwidth]{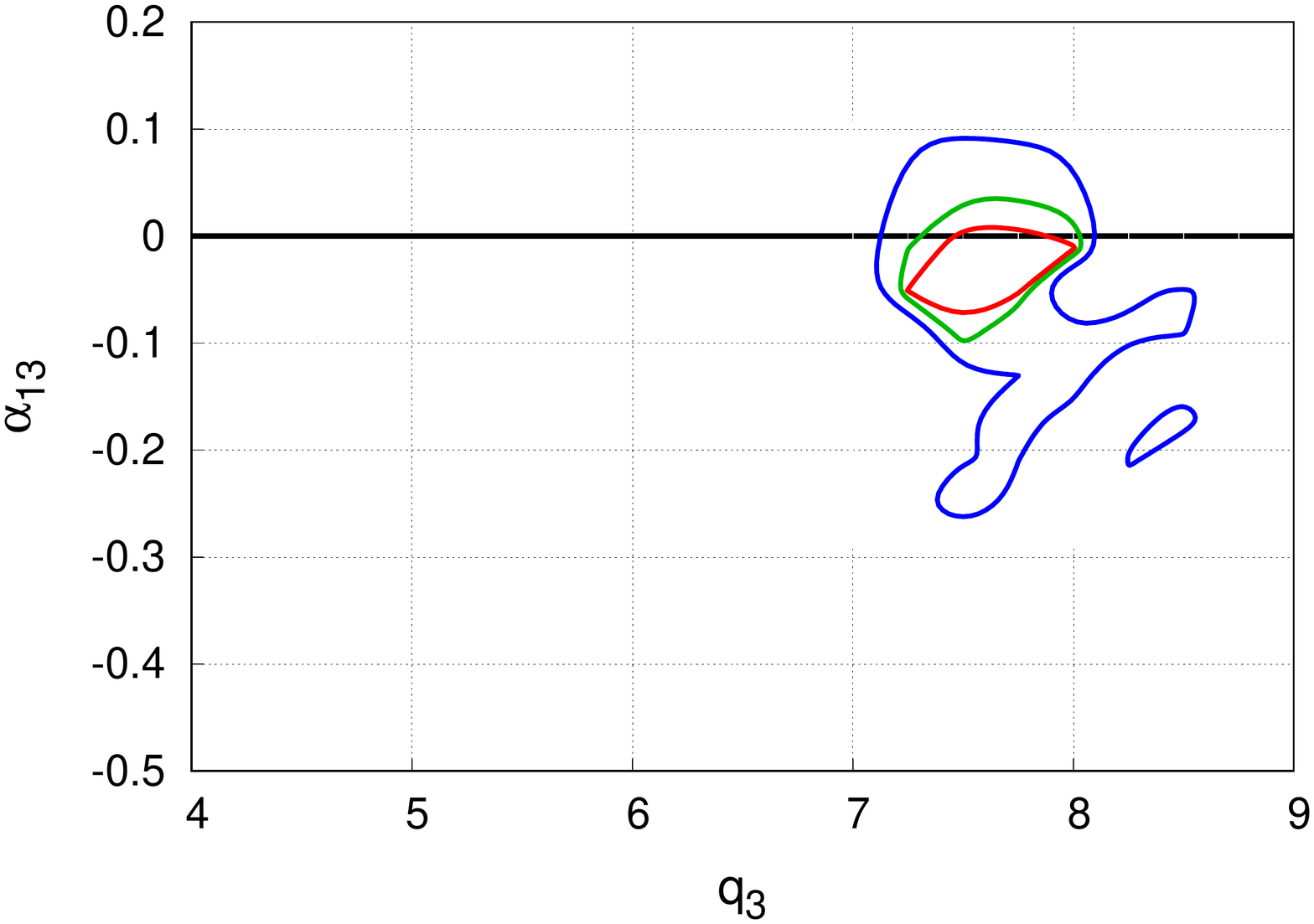}
\hspace{-0.5cm}
\includegraphics[type=pdf,ext=.pdf,read=.pdf,width=0.5\textwidth]{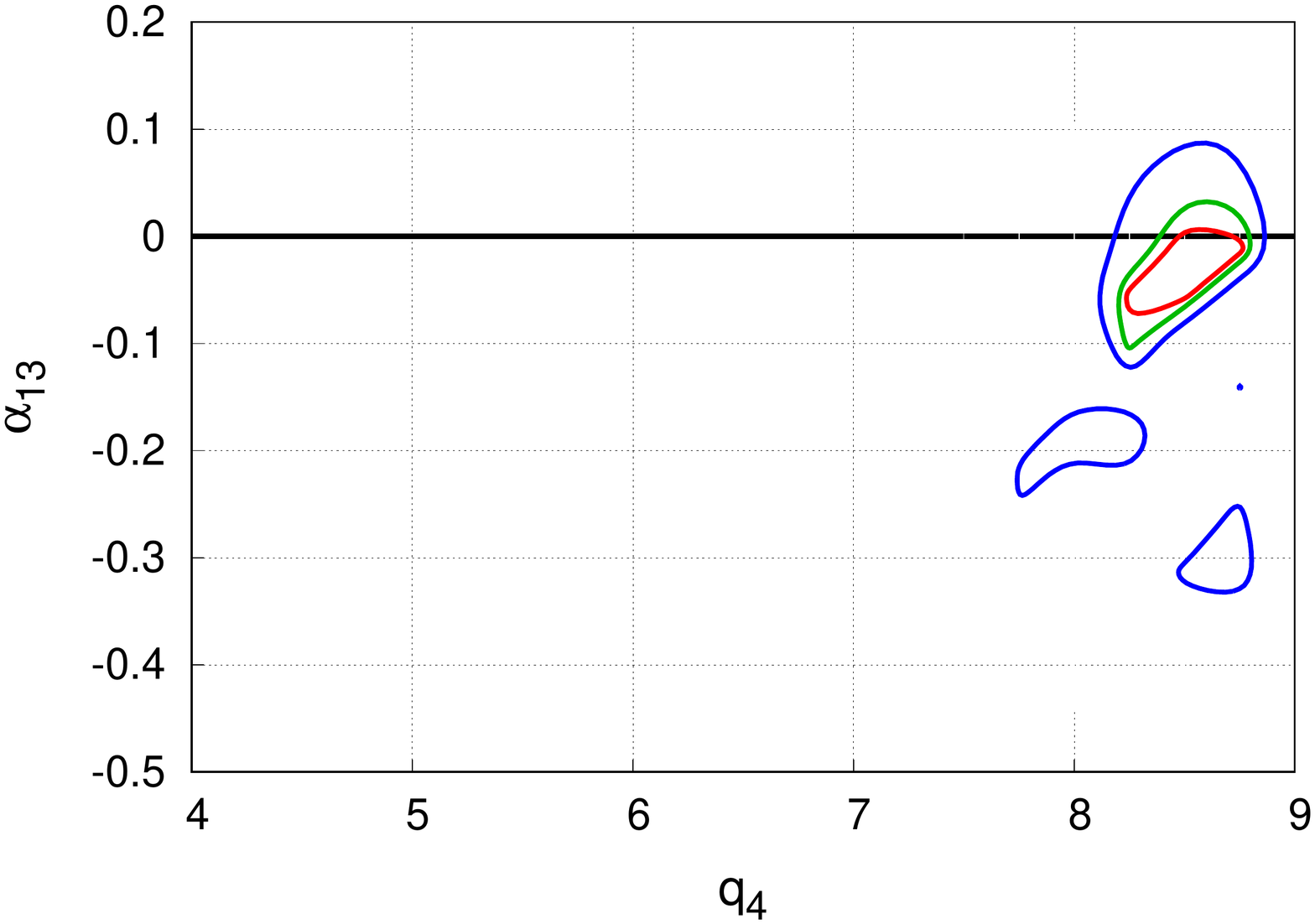} \\
\end{center}
\vspace{-1.2cm}
\caption{Impact on deformation parameter $\alpha_{13}$ by emissivity profile of different flux intervals.
$q_1$, $q_2$, $q_3$, and $q_4$ are the inner emissivity indices for low, medium, high and very high flux state, respectively. The red, green, and blue lines indicate, respectively, the 68\%, 90\%, and 99\% confidence level contours for two relevant parameters. The thick black horizontal line marks the Kerr solution. \label{f-qn}}
\end{figure*}


\section{Discussion and conclusions \label{s-dis}}

The primary aim of this work is put constraints on possible deviations from the Kerr solution. For $\alpha_{22} = 0$, our constraints on the black hole spin parameter $a_*$ and the Johannsen deformation parameter $\alpha_{13}$ are (99\% confidence level for two relevant parameters)
\be
0.928 < a_* < 0.983 \, , \qquad
-0.44 < \alpha_{13} < 0.15 \, .
\ee
When we assume $\alpha_{13} = 0$, we find (still 99\% confidence level for two relevant parameters)
\be
0.885 < a_* < 0.987 \, , \qquad
-0.12 < \alpha_{22} < 1.05 \, .
\ee
In both cases, the value of the deformation parameter is consistent with 0, which is the value required by the Kerr solution and predicted by Einstein's gravity. The confidence level contours $a_*$ vs $\alpha_{13}$ and $a_*$ vs $\alpha_{22}$ are shown in Fig.~\ref{f-plots}.

The best-fit values of $\alpha_{13}$ and $\alpha_{22}$ are very close to 0, so it is relatively straightforward to compare the results obtained here with those from previous studies. In general, we can say that the best-fit values of the model parameters are consistent with the estimates found in previous analyses in the Kerr background, and in particular with those found in~\cite{s9} analyzing the same \textsl{XMM-Newton} and \textsl{NuSTAR} observations. For the model parameters that are constant over different flux states, our measurements are consistent with those in~\cite{s9}. The estimate of the spin parameter and of the inclination angle of the accretion disk agree with those in~\cite{s9}. All previous studies find for this source an iron abundance higher than the Solar iron abundance~\cite{s5}. The iron abundance found in~\cite{s9} is somewhat lower than ours, but the difference can be easily explained by the different analysis method. For the model parameters that vary over different flux states, a direct comparison with Ref.~\cite{s9} is not possible because the grouping scheme is different from ours. We just notice that here we find that the ionization parameter of the relativistic reflection component nicely increases with the luminosity, as is expected.

\subsection{Constraints on the Kerr metric}

As in our previous studies, our results are consistent with the Kerr black hole hypothesis. The constraint on $\alpha_{13}$ obtained in the present work from MCG--06--30--15 is comparable to that obtained from the stellar-mass black hole in GS~1354--645~\cite{noi5}, while the constraint on $\alpha_{22}$ is slightly weaker, see Eqs.~(\ref{eq-a13-gs}) and (\ref{eq-a22-gs}).

On one hand, the fact that we mostly recover the Kerr solution is expected: since Einstein's gravity has already successfully passed a large number of observational tests, it is likely that astrophysical black holes are at least very similar to, if not exactly, the Kerr black holes of Einstein's gravity. On the other hand, since our model has a number of simplifications that inevitably introduce many systematic uncertainties, these confirmations are not at all obvious. It could be that our non-Kerr parameters were able to absorb some of these systematic uncertainties. We are thus tempted to argue that the fact we always recover Kerr can be interpreted as the systematic uncertainties from the model being currently subdominant relative to the uncertainties due to the quality of the data.

Model simplifications are both in the description of the accretion disk and in the calculations of the reflection spectrum. The disk is assumed to be infinitesimally thin, on the equatorial plane of the black hole, and its inner edge is set at the innermost stable circular orbit (ISCO). In the reality, the thickness of the disk is finite and increases with the mass accretion rate. A preliminary study on the impact of the disk thickness on the reflection spectrum has been reported in~\cite{taylor}. The inner edge of the disk is thought to be at the ISCO radius when the accretion luminosity is between 5\% and 30\% of the Eddington limit~\cite{isco1,isco2}, while for higher luminosities it may move to a smaller radius~\cite{isco3}. For supermassive black holes, it is typically difficult to get reliable estimates of the accretion luminosity, because of the large uncertainties in the estimates of their mass and distance from us. In the case of MCG--06--30--15, the Eddington scaled accretion luminosity has been estimated to be $0.40 \pm 0.13$~\cite{brenneman}, so deviations from the thin disk model can be expected even if they may be moderate.

There are also a number of simplifications in the calculation of the reflection spectrum. Our model currently assumes a fixed electron density in the disk, a constant disk density over height and radius, a single ionization parameter for the whole disk, etc. At the moment, a systematic study on the impact of these simplifications on the measurements of the spin and the deformation parameters is lacking, but work is underway.

The emissivity profile is usually thought to be a crucial ingredient and source of systematic uncertainties. To check its impact on the estimate of the deformation parameters, in~\cite{noi5} we showed that incorrect modeling of the emissivity profile leads to non-vanishing deformation parameters. The fact that we always recover the Kerr metric when we fit the emissivity index suggests that the quality of our data is good enough to permit us to estimate both the deformation parameter and the emissivity index, as an accidental compensation leading to recovery of the Kerr metric sounds unlikely. However, if the emissivity profile is so important for the estimate of the deformation parameter, as suggested in~\cite{noi5}, we should expect that a power law or a broken power law may not be adequate in the case of high quality data, hopefully available with the next generation of X-ray missions~\cite{extp-snz}.

To further explore the role of the emissivity profile on the estimate of the deformation parameters, we have plotted the constraints on the plane $q_{\rm in}$ vs $\alpha_{13}$ for every flux state of the observations of MCG--6--30--15 studied in this work. Fig.~\ref{f-qn} shows the 68\%, 90\%, and 99\% confidence level contours for two relevant parameters, where $q_1$, $q_2$, $q_3$, and $q_4$ are, respectively, the inner emissivity indices for low, medium, high and very high flux state. These plots do not show any particular correlation between the $q_{\rm in}$ and $\alpha_{13}$, confirming that the spectral analysis of the source can separately determine these two quantities. Such a conclusion cannot, in general, be extended to any deformation parameter, but at least it seems to hold for $\alpha_{13}$ and $\alpha_{22}$. Because of the current uncertainties in the corona geometry, and therefore in the exact shape of the emissivity profile, the non-observation of a correlation between the emissivity index and the deformation parameters can partially limit the systematic uncertainty in the estimate of the deformation parameters due to the uncertainty in the correct emissivity profile.


{\bf Acknowledgments --}
A.T. thanks Laura Brenneman for useful discussions on MCG--06--30--15. This work was supported by the National Natural Science Foundation of China (NSFC), Grant No.~U1531117, and Fudan University, Grant No.~IDH1512060. A.T. also acknowledges support from the China Scholarship Council (CSC), Grant No.~2016GXZR89. S.N. acknowledges support from the Excellence Initiative at Eberhard-Karls Universit\"at T\"ubingen. A.B.A. also acknowledges the support from the Shanghai Government Scholarship (SGS). J.A.G. acknowledges support from the Alexander von Humboldt Foundation.


\end{document}